\documentclass[fleqn,usenatbib]{mnras}
\usepackage{mathptmx}
\usepackage{txfonts}

\usepackage[utf8]{inputenc}
\usepackage[T1]{fontenc}

\DeclareRobustCommand{\VAN}[3]{#2}
\let\VANthebibliography\thebibliography
\def\thebibliography{\DeclareRobustCommand{\VAN}[3]{##3}\VANthebibliography}

\usepackage{savesymbol}	% Including figure files
\savesymbol{negmedspace}
\savesymbol{negthickspace}
\savesymbol{iint}
\savesymbol{iiint}
\savesymbol{iiiint}
\savesymbol{idotsint}
\usepackage{graphicx}	% Including figure files
\usepackage{amsmath}	% Advanced maths commands
\usepackage{amssymb}	% Extra maths symbols
\usepackage{xcolor}
\definecolor{grey}{rgb}{0.4,0.4,0.4}
\definecolor{dred}{rgb}{1.0,0.0,0.15}

\title[Rossby numbers of convective stars]{Rossby numbers of fully and partially convective stars}

\author[N. R. Landin et al.]{
N. R. Landin,$^{1,2}$\thanks{E-mail: nlandin@ufv.br}
L. T. S. Mendes,$^{3,2}$
L. P. R. Vaz$^{2}$
and S. H. P. Alencar$^{2}$
\\
$^{1}$Universidade Federal de Vi\c cosa, Campus UFV Florestal, CEP 35690-000 -- Florestal, MG, Brazil\\
$^{2}$Depto.\ de F\'{\i}sica, Universidade Federal de Minas Gerais, C.P.702, 31270-901 -- Belo Horizonte, MG, Brazil\\
$^{3}$Depto.\ de Engenharia Eletr\^onica, Universidade Federal de Minas Gerais, C.P.702, 31270-901 -- Belo Horizonte, MG, Brazil
}

\date{Accepted XXX. Received YYY; in original form ZZZ}

\pubyear{2022}

\begin{document}
\label{firstpage}
\pagerange{\pageref{firstpage}--\pageref{lastpage}}
\maketitle

\begin{abstract}
We investigate stellar magnetic activity from the theoretical point of view, by using
stellar evolution models to calculate theoretical convective turnover times
($\tau_{\rm c}$) and Rossby numbers (${\rm Ro}$) for pre-main-sequence and
main-sequence stars. The problem is that the canonical place where $\tau_{\rm c}$ is
usually determined (half a mixing length above the base of the convective zone) fails
for fully convective stars and there is no agreement on this in the literature. Our
calculations were performed with the {\ttfamily ATON} stellar evolution code. We
concentrated our analysis on fully and partially convective stars motivated by recent
observations of slowly rotating fully convective stars, whose X-ray emissions
correlate with their Rossby numbers in the same way as in solar-like stars,
suggesting that the presence of a tachocline is not required for magnetic field
generation. We investigate the behaviour of $\tau_{\rm c}$ over the stellar radius
for stars of different masses and ages. As ${\rm Ro}$ depends on $\tau_{\rm c}$,
which varies strongly with the stellar radius, we use our theoretical results to
determine a better radial position at which to calculate it for fully convective
stars.  Using our alternative locations, we fit a sample of 847 stars in the
rotation-activity diagram ($L_{\rm X}/L_{\rm bol}$\,{\it versus}\,${\rm Ro}$) with a
two-part power-law function.  Our fit parameters are consistent with previous work,
showing that stars with ${\rm Ro}$$\leq$${\rm Ro_{sat}}$ are distributed around a
saturation level in $L_{\rm X}/L_{\rm bol}$ and, for stars with ${\rm Ro}$$>$${\rm
Ro_{sat}}$, $L_{\rm X}/L_{\rm bol}$ clearly decays with ${\rm Ro}$ with an exponent
of $-2.4\!\pm\!0.1$. 
\end{abstract}


%%%%%%%%%%%%%%%%%%%%%%%%%%%%%%%%%%%%%%%%%%%%%%%%%%

%%%%%%%%%%%%%%%%% BODY OF PAPER %%%%%%%%%%%%%%%%%%

\section{Introduction}

Stars of different spectral types and ages host large-scale magnetic fields, as
evidenced by observable phenomena like star spots, flares, activity cycles, coronal
heating and chromospheric and coronal emissions, all of which express in some way
stellar magnetic activity. Magnetic fields are observed in most (if not all) low-mass
pre-main-sequence (pre-MS) and main-sequence (MS) stars \citep[spectral types F, G, K
and M,][]{donati09}. However, only 5-10 per cent of intermediate mass stars (Herbig
Ae/Be and A-B5 MS stars, \citealt{villebrun19} and \citealt{alecian19}) and 7\%\ of
high-mass MS stars \citep[of spectral types O-B5,][]{keszthelyi20} exhibit detectable
magnetic fields.

From a theoretical point of view, a moving plasma subjected to a magnetic field is
described by the magneto-hydrodynamic (MHD) equations and the dynamo theory is used
as a main tool to investigate the processes responsible for keeping and regenerating
stellar magnetic fields. The mechanism driving magnetic activity is generally
attributed to a dynamo resulting from the interaction between differential rotation
and convective motions in the outer convective envelope of the star. Theoretical
results point out that, for MS solar type stars, the magnetic field is generated and
amplified at the tachocline, a thin layer of differential rotation located at the
interface between the internal radiative core and the external convective envelope.
For stars of spectral types between F and M, magnetic activity and rotation are
believed to be regulated by a dynamo process called $\alpha$-$\Omega$ effect
\citep{mohanty03}. In the $\Omega$ effect, differential rotation distorts the
poloidal field in order to generate the toroidal field, while in the $\alpha$ effect,
helical turbulence twists the toroidal field to regenerate the poloidal field. In
this type of dynamo, the $\alpha$ and $\Omega$ effects are such that the poloidal and
toroidal field components sustain themselves through a cyclic feedback process
\citep{nelson08}. The efficiency of this process depends on the rotation rate and on
the convective motion time scale.  Young, fast rotating stars are generally very
active.  Dynamo theory models, such as the $\alpha$-$\Omega$ (or interface dynamo),
have been successful in explaining qualitative features of solar magnetic activity
\citep{weiss00}. For fully convective stars, such as very low-mass MS stars
($M$$<$0.35\,M$_{\odot}$) and very young stars in the pre-MS, this theory cannot be
applied, because they are missing a tachocline. However, there are observations of
magnetic activity indicators of fully convective stars and measurements of their
magnetic field topology \citep{morin10}, some kind of dynamo should be in action in
these stars. \citet{durney93} suggested that a distributed or turbulent dynamo could
be operating in completely convective stars.

From the observational point of view, \citet{kraft67}, \citet{frazier70} and
\citet{skumanich72} carried out pioneering works relating rotation to magnetic
activity. \citet{skumanich72} suggested for the first time that the rotation-activity
relationship was a consequence of the dynamo action and showed that the stellar
rotation rate decreases with age as $t^{-1/2}$.  \citet{pallavicini81} showed that
the fractional X-ray luminosity ($L_{\rm X}/L_{\rm bol}$) is correlated with the
projected rotational velocity $\varv\sin i$ (and anti-correlated with rotation
period). However, as pointed out by \citet{noyes84}, the rotation-activity
correlation is usually better described in terms of the Rossby number ${\rm Ro}$,
defined as the ratio of the rotational period $P_{\rm rot}$ to the local convective
turnover time $\tau_{\rm c}$ (${\rm Ro}$=$P_{\rm rot}/\tau_{\rm c}$).  By analysing
chromospheric activity indicators in stars of spectral types between F and M,
\citet{soderblom93} verified that chromospheric magnetic activity increases rapidly
with rotational velocity until $\varv\sin i$$\approx$${\rm 15~km/s}$, and then
reaches a saturation plateau above this velocity threshold.  Similar behaviours are
also found with coronal magnetic activity indicators, as shown by, e.g.,
\citet{stauffer94}.  When analysed in terms of Rossby numbers, the saturation occurs
for ${\rm Ro}$$\lesssim$0.1, which is equivalent to rotation periods of 1\,to\,10\,d
for solar-like stars.  \citet{pizzolato03} extended the sample of \citet{noyes84} and
showed the existence of two distinct regions: the saturated region (formed by fast
rotators) and the unsaturated region (composed by slow rotators).  \citet{prosser96}
suggested the existence of a third regime, called supersaturation, which occurs for
small values of ${\rm Ro}$ (${\rm Ro}$$<$0.01). In this regime, also observed by
\citet{randich96}, stars with high rotation rates exhibit magnetic activity levels
below the saturation level. According to \citet{argiroffi16}, two physical mechanisms
can be considered to explain this phenomena: centrifugal stripping\footnote{%
A predicted reduction of the coronal emission at high rotational velocities, due to
the disruption of the largest coronal structures by centrifugal forces.}
\citep{jardine99} and polar updraft migration\footnote{%
Polar updraft migration is that, at high rotational velocities, the magnetic flux
emergence becomes more efficient near the poles, diminishing or even vanishing
magnetic flux tubes and coronal structures at equatorial regions.}
\citep{stepien01}.

Magnetic activity indicators can be expressed as a function of ${\rm Ro}$ in the
unsaturated region, but they are constant in the saturated regime.  It characterises
the rotation-activity relationship, which is a powerful tool in studies involving
stellar dynamo processes, based on a general power-law function of the form $L_{\rm
X}/L_{\rm bol}$$\propto$${\rm Ro}^{\beta}$ (using $L_{\rm X}/L_{\rm bol}$ as an
indicator of coronal activity), where $\beta$ is a parameter to be adjusted with
observations and canonically is $-2$ \citep{pizzolato03}.  More recently,
\citet{wright11}, \citet{wright18} and \citet{newton17} analysed the
rotation-activity relationship for samples of 824, 847 and 466 stars, respectively.
The first two determined the level of the saturation regime in $L_{\rm X}\!\! \approx
\!\! 10^{-3} L_{\rm bol}$ and power-law slopes of $\beta \!\! = \!\! -2.70\! \pm \!
0.13$ \citep{wright11} and $\beta \!\! = \!\! -2.3^{+0.4}_{-0.6}$ \citep{wright18}.
\citet{newton17} found $\beta=-1.7{\pm}0.1$ and a saturation level of $L_{\rm
H_{\alpha}}\!\! \approx \!\! 10^{-4}L_{\rm bol}$, where $L_{\rm H_{\alpha}}$
($H_{\alpha}$ luminosity) is a chromospheric activity indicator.

The Rossby number is an important parameter not only in stellar activity
investigations, but also in astrophysical fluid dynamics (${\rm Ro}$ is a measure of
the importance of rotation in the fluid flow) and in MHD simulations. By describing
the rotation-activity relationship in terms of the Rossby number, one can readily see
its connection with stellar dynamo models.  The reason for this, as pointed out by
\citet{kim96}, is that the dynamo number $N_{\rm D}$, a dimensionless parameter in
the mean-field dynamo theory which characterises the model behaviour, is proportional
to the inverse square of Rossby number.  By its turn, the dynamo number is
essentially the ratio between the terms of magnetic field generation and diffusion.
As long as $N_{\rm D}\propto {\rm Ro}^{-2}$, the dynamo efficiency, and consequently
the level of magnetic activity, increases with decreasing Rossby number.  As ${\rm
Ro}$ plays an important role in the stellar magnetic activity studies, determinations
of $\tau_{\rm c}$ are of fundamental interest, because they cannot be directly
observed. Unfortunately, our limited knowledge of stellar convection imposes a strong
challenge for correct convective turnover time calculations. Even when we rely on the
Mixing Length Theory approximation, extensively used in the literature, some
uncertainties are still involved. For instance, the assumption of the same mixing
length parameter $\alpha$ for modelling stars of different masses and evolutionary
stages \citep{kim96}.

Though $\tau_{\rm c}$ can only be assessed theoretically with stellar models, it has
been customary in the literature to employ its semi-empirical value, a theoretically
derived $\tau_{\rm c}$ expressed as function of an observed colour index.  In their
works concerning the rotation-activity relation, \citet{noyes84} and
\citet{pizzolato03} used $\tau_{\rm c}(B\!\!-\!\!V)$, \citet{wright18} utilised
$\tau_{\rm c}(V\!\!-\!\!K)$ and \citet{wright11} handle with both semi-empirical
colour functions of $\tau_{\rm c}$.  Among the published estimates of $\tau_{\rm c}$
which come from models that track the stellar evolution, we describe four in more
detail.  \citet{gilliland86} did theoretical estimates of convective turnover times
by using simple standard models described by \citet{eggleton71,eggleton72}.
\citet{kim96} published convective turnover times for solar-like stars in the pre-MS
and early post-main-sequence phases of evolution. They used non-grey boundary
conditions, but the opacities used in their work are less up-to-date than those we
use.  \citet{jung07} presented convective turnover times for low- and very low-mass
stars in the pre-MS. Although they published a finer grid resolution than we do,
their models did not take into account stellar rotation and used grey boundary
conditions, which are not suitable for effective temperatures below about 4000K
\citep{baraffe97}.  \citet{landin10} computed convective turnover times for rotating,
grey models ranging from 0.6 to 1.2\,M$_\odot$ with solar chemical composition.

The rotation-activity relationship (magnetic activity increasing with decreasing
Rossby number down to ${\rm Ro}\approx0.1$ and remaining constant for ${\rm
Ro}\leq0.1$), clearly observed for MS stars, cannot be seen among pre-MS stars (e.g.
\citealt{flaccomio03} and \citealt{feigelson03}).  In this evolutionary phase, all
stars are found to be in the saturated regime and a considerable dispersion in the
magnetic activity levels is observed, as \citet{preibisch05}, \citet{alexander12} and
\citet{argiroffi14} have shown for stars in Orion Nebula Cluster (1\,Myr), IC348
(2-3\,Myr) and h~Per (13\,Myr), respectively.  Until recently, it was believed that
the saturated part of the rotation-activity relationship was composed by partially
and totally convective stars with high rotation rates while the unsaturated part was
formed only by partially convective stars with slow rotation rates.  However, recent
observations by \citet{wright16} and \citet{wright18} of X-ray emission of slowly
rotating fully convective stars indicate that partially and fully convective stars
follow the same rotation-activity relationship. According to \citet{wright16}, as
this relationship is considered a proxy for the behaviour of the magnetic dynamo,
these results imply that partially and fully convective stars should operate very
similar rotation-dependent dynamos.  Because standard dynamo model is based on the
shearing of internal magnetic fields by differential rotation that takes place at the
tachocline, which is not present in fully convective stars, the results of
\citet{wright16} and \citet{wright18} imply that the presence of a tachocline is not
a central key for magnetic field generation.  In these cases, as they also mentioned,
there are other possibilities for large-scale magnetic field generation such as the
turbulent or the distributed dynamos \citep[see e.g.][for a review on this
subject]{brun17}.

In order to contribute to a better understanding of the dynamo mechanism existing in
fully and partially convective stars, we study the behaviour of the parameters
related to Rossby numbers inside stars of different masses and ages with the
{\ttfamily ATON} stellar evolution code.  For a given stellar mass and age, the
convective turnover time changes significantly depending on the location inside the
star at which it is calculated. The location mostly used in the literature to
determine $\tau_{\rm c}$ is near the base of the convective zone, one mixing length
\citep{gilman80} or one half a mixing length \citep{noyes84} above it.  This choice
is based on the assumption that the source region of the stellar magnetic dynamo
matches that of the tachocline \citep{parker75}.  Observational results showing that
the intensity of the chromospheric emission (H \& K lines of Ca II) of solar-type
stars and Rossby number calculated at one half a mixing length above the base of the
convective zone correlate well \citep{schatzman93}, seem to support it.  However,
these locations are not suitable for fully convective stars, including young pre-MS
stars and MS very low-mass stars. 

Section~\ref{models} briefly presents the physical ingredients of the  {\ttfamily
ATON} code and the input parameters used in this work.  Section~\ref{calculations}
presents our theoretical convective turnover times and convective velocities.  Aiming
at finding an alternative location to calculate $\tau_{\rm c}$ for young stars and
very low-mass stars, but still related to the default location, we evaluate
$\tau_{\rm c}$ throughout the whole star, investigate its behaviour for different
stellar masses and ages, analyse where it is usually calculated for masses greater
than 0.4\,M$_{\odot}$ (mass interval in which all stars are partially convective in
the MS), and propose an alternative location to calculate it for fully convective
stars.  Section~\ref{results} discusses results on local convective turnover times
and Rossby number calculations.  In Section~\ref{applications}, our new theoretical
convective turnover times are compared with those available in the literature and
observational data from low-mass and very-low mass stars are used to test our
theoretical results. Finally, our conclusions are given in
Section~\ref{conclusions}.

\section{Models and input physics} \label{models}

In the {\ttfamily ATON} code version used in this work, convection is treated
according to the traditional Mixing Length Theory \citep{bohmvitense58}, with the
parameter that represents the convection efficiency $\alpha$$=$$2$. Surface boundary
conditions were obtained from non-grey atmosphere models \citep{allard00} with
matching between the surface and interior at optical depth $\tau\!\!=\!\!3$.  We used
the opacities reported by \citet{rogers1} and \citet{alexander} and the equations of
state from \citet{rogers2} and \citet{mihalas}.  Here, we assume that the elements
are mixed instantaneously in convective regions.  Our tracks start from a fully
convective configuration with central temperatures in the range $5.35<\log_{10}
T_{\rm c}<5.72$, follow deuterium and lithium burning, and end at the MS
configuration. For a  discussion about the zero point of ages of stellar models see
\citet{landin06}.

The current version of the {\ttfamily ATON} code allows choosing among three
rotational schemes \citep{mendes99}, (1) rigid body rotation throughout the whole
star, (2) local conservation of angular momentum in the whole star (which leads to
differential rotation) and (3) local conservation of angular momentum in radiative
regions plus rigid body rotation in convective regions. However, there is
observational evidence that the Sun's radiative core rotates as a solid body and the
convective envelope rotates differentially, opposite to scheme 3 \citep{thompson03}.
Here, our rotating models were generated according to scheme 2 because differential
rotation in the convective envelope is an important ingredient to the dynamo process.
We leave for the future implementation of a 4$^{th}$ rotational scheme with a
rotational profile closer to that of the Sun.

The initial angular momentum of each model is obtained according to \citet{kawaler87}
as
\begin{equation}
J_{\rm kaw}=1.566 \times 10^{50} \left( {M \over M_{\odot}} \right) ^{0.985}
~~~{\mathrm{g~cm^2~s^{-1}.}}
\label{eqkaw}
\end{equation}

The evolutionary tracks were computed in the mass range of 0.1 to 1.5\,M$_{\odot}$
(in 0.1\,M$_{\odot}$ steps). We adopt the solar chemical composition $X$=$0.7155$ and
$Z$=$0.0142$, by \citet{asplund09}.  More details on the physics of the {\ttfamily
ATON} models are given by \citet{landin06}.

\section{Convective turnover times calculations} \label{calculations}

There are two ways to calculate theoretical convective turnover times using stellar
evolution models, locally and non-locally.  This section presents the details of such
calculations in the {\ttfamily ATON} code. We also investigate how convective
turnover times and convective velocities vary with stellar mass and age. For some
selected ages, we examine the behaviour of radial profiles of these quantities for
stellar masses from 0.1 to 1.5\,M$_{\odot}$, aiming to find an alternative location
to calculate convective turnover times locally.

\begin{figure} %Fig 1
\centering{
\includegraphics[width=\columnwidth]{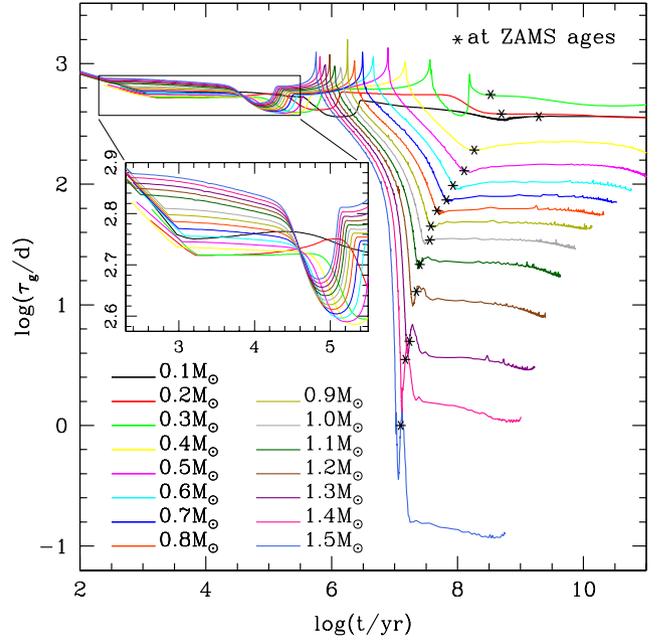}
\caption{Global convective turnover time as a function of age $t$ and stellar mass.
Asterisks (*) show the ZAMS ages for each mass model.  The insert shows in detail the
temporal evolution of $\tau_{\rm g}$ during the beginning of the pre-MS phase.
Throughout this work all the logarithms are taken in base 10.}
\label{taugxage}
}
\end{figure}

\subsection{Global convective turnover times}\label{taug}

Following \citet{kim96}, we evaluate the global (or non-local) convective turnover
time ($\tau_{\rm g}$) for each model (see Fig.\,\ref{taugxage}). The global
convective turnover time is defined as $\tau_{\rm g}\!\!=\!\!\! \int_{R_{\rm
b}}^{R_{\rm star}}\varv_{\rm c}^{-1}{\rm d}r$, where $R_{\rm b}$ is the radial
position of the base of the convective zone relative to the centre of the star,
$R_{\rm star}$ is the stellar radius and $\varv_{\rm c}$ is the convective velocity.
According to \citet{kim96}, this is the characteristic time-scale for convective
overturn. It can be used to describe convection features of the whole stellar
convective region at each evolutionary stage.  We note that in Fig.\,\ref{taugxage}
the time evolution of $\tau_{\rm g}$ differs for fully and partially convective
stars.  Global convective turnover times of very low-mass stars
($M\!\!\le$0.3\,M$_{\odot}$) present small variations (less than 0.6\,dex -- a factor
of 3.5), while for stars with $M\!\!>$0.3\,M$_{\odot}$ variations in $\tau_{\rm g}$
can reach almost 4 orders of magnitude.  Values of $\tau_{\rm g}$ do not change
significantly near the beginning of the pre-MS phase, in which stars are fully
convective, decrease when stars develop a radiative core and remain roughly constant
during the MS phase. For $M\!\!>$1.3\,M$_{\odot}$, $\tau_{\rm g}$ experiences an
extra reduction after the zero-age main sequence (ZAMS, asterisks in
Fig.\,\ref{taugxage}) and before reaching its nearly constant MS value.  It reflects
the behaviour of the pressure scale-height for such models.  During the pre-MS, the
higher the stellar mass the higher $\tau_{\rm g}$, except for $M$$\le$0.3\,${\rm
M_{\odot}}$, when the opposite situation is observed (see the amplified view in
Fig.\,\ref{taugxage}).  On the MS, $\tau_{\rm g}$ decreases as the mass increases for
$M$$>$0.3$\,{\rm M_{\odot}}$ but for $M$$\le$0.3\,${\rm M_{\odot}}$ $\tau_{\rm g}$
increases with mass.

\begin{figure} %Fig 2
\centering{
\includegraphics[width=\columnwidth]{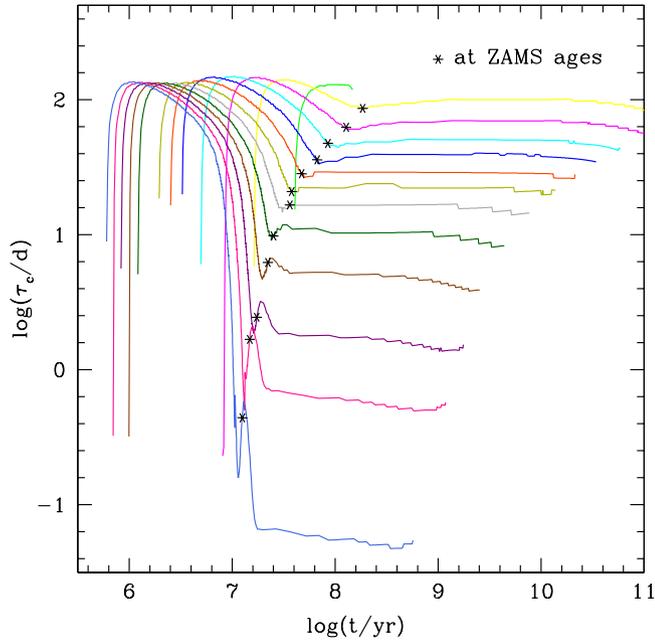}
\caption{Local convective turnover time as a function of age and stellar mass.
Symbols and colours have the same meanings as in Fig.\,\ref{taugxage}. The
calculations were performed only while $r_{\mathrm{std}}$$<$$R_{\mathrm{star}}$,
where $r_{\mathrm{std}}$, our standard location to calculate $\tau_{\rm c}$, is
defined as one-half of a mixing length above the base of the convective zone.}
\label{taucxage}
}
\end{figure}

\subsection{Local convective turnover times}\label{tauc}

For the purpose of computing Rossby numbers, the characteristic time-scale used is
the local convective turnover time, calculated in the deep convection envelope where
dynamo generation of magnetic fields is supposed to take place.  The local convective
turnover time is defined as the ratio of convection mixing length ($\ell$) to
convective velocity, $\tau_{\rm c}\!\!=\!\!\ell/\varv_{\rm c}$.  Following
\citet{noyes84}, we adopt as the standard location ($r_{\rm std}$) to calculate
$\tau_{\rm c}$ at one-half of a mixing length above the base of the convective zone
$r_{\rm std}\!\!=\!\!R_{\rm b}\!+\!\ell/2$, where $\ell$ is evaluated at the base of
the convective zone.  In Fig.\,\ref{taucxage}, we show local convective turnover
times calculated at the standard location as a function of age, for some stellar
models.  Local convective turnover times vary significantly with age before the ZAMS
and remain virtually constant during the MS.  On the MS $\tau_{\rm c}$ decreases with
the stellar mass.  One can notice in Fig.\,\ref{taucxage} that the
0.3\,${\rm M_{\odot}}$ model (green curve) seems to be incomplete and the 0.1 and
0.2\,${\rm M_{\odot}}$ models (black and red curves, respectively) are not shown.
The reason is that, for fully convective stars, the base of the convective zone is
the stellar centre, which is singular \citep{gilliland86}. Besides $M(r)=0$ and
$L(r)=0$ at the stellar centre, in Mixing Length Theory, the adjustable parameter,
the mixing length, is scaled with the local pressure scale height $H_{\rm p}$ through
the relation $\ell\!=\!\alpha H_{\rm p}$ and, for these models, $H_{\rm p}$ at the
base of the convective zone is very high, and so is the mixing length.  Consequently,
the place where $\tau_{\rm c}$ should be calculated,
$r_{\rm std}\!=\!R_{\rm b}\!+\!\alpha H_p/2$, becomes larger than the stellar radius.
As $r_{\rm std}$ depends on the adjustable parameter $\alpha$, one could think of
using an alternative $\alpha_{\rm alt}$ whenever $r_{\rm std}\!\!>\!\!R_{\rm star}$.
However, to accomplish this, $\alpha_{\rm alt}$ should be smaller than 0.02, at least
two orders of magnitude smaller than that which reproduces the Sun, implying a very
inefficient convective energy transport.  So, these facts reinforce the idea that
applying prescriptions based explicitly on the tachocline to obtain the place where
the magnetic field is produced and amplified in fully convective stars is senseless
\citep{feiden13}.  For all the reasons discussed above, we believe that the usual
location in the stellar interior used to calculate ${\rm Ro}$ is not suitable for
fully convective stars. In cases which $r_{\rm std}\!\!>\!\!R_{\rm star}$,
$\tau_{\rm c}$ is not calculated.

\begin{figure} %Fig 3
\centering{
\includegraphics[width=\columnwidth]{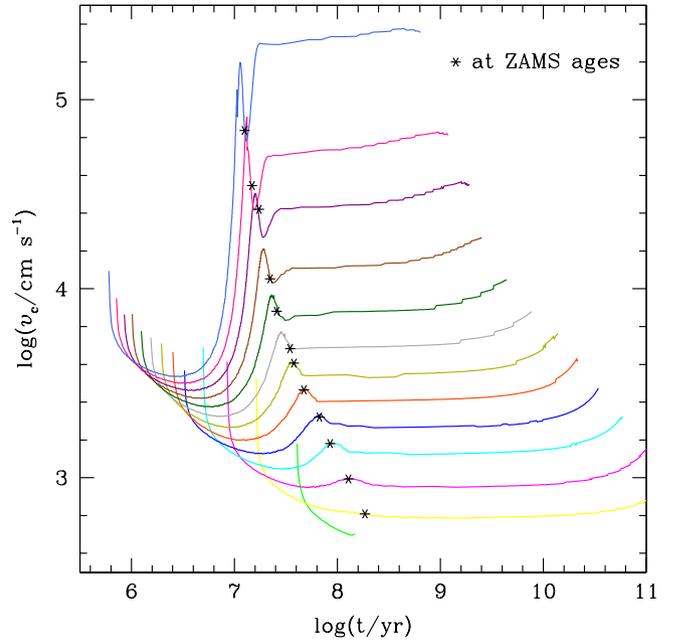}
\caption{Convective velocity as a function of age and stellar mass.
Symbols and colours have the same meanings as in Fig.\,\ref{taugxage}. The
calculations were performed only while $r_{\mathrm{std}}$$<$$R_{\mathrm{star}}$.}
\label{velcxage}
}
\end{figure}

\subsection{Convective velocities}\label{velc}

Both $\tau_{\rm c}$ and $\tau_{\rm g}$ represent characteristic time scales for
stellar convection and so should differ only by a constant factor.  Because these two
quantities depend on the convective velocity, we show how this behaves as function of
stellar age and mass in Fig.\,\ref{velcxage}, again only for ages and masses for
which $\tau_{\rm c}$ can be calculated at the standard location. Before reaching the
ZAMS (marked with asterisks), $\varv_{\rm c}$ varies significantly with stellar age
but remains roughly constant on the MS. For models with $M\!>\!$1.3\,M$_{\odot}$,
$\varv_{\rm c}$ is subjected to an extra variation before becoming approximately
constant on the MS.  This is due to variations in the pressure scale-height occurring
in such models.  In this phase, the convective velocity increases with the stellar
mass.

\subsection{Radial profiles of ${\rm Ro}$, $\tau_{\rm c}$ and $\varv_{\rm c}$ at the ZAMS}\label{radprof}

\begin{figure} %Fig 4
\centering{
\includegraphics[width=\columnwidth]{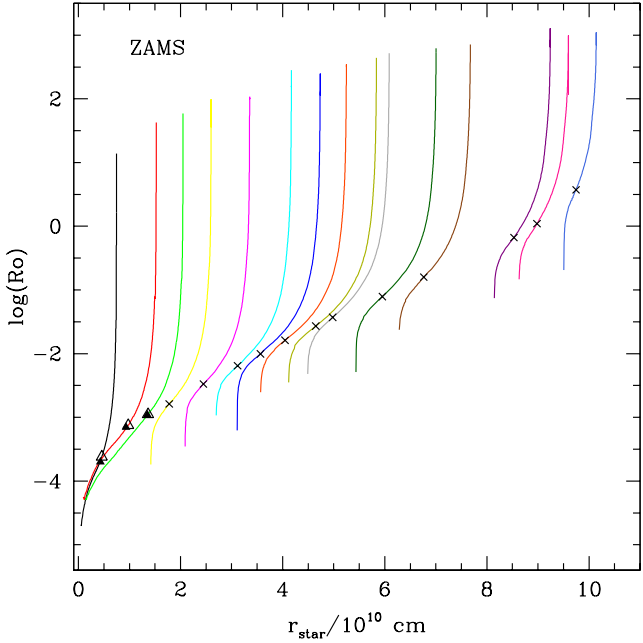}
\caption{Rossby number Ro as a function of radius $r_{\mathrm{star}}$ through the
star for each mass model.  Crosses ($\times$) show the standard locations inside the
stars where $\tau_{\rm c}$ is usually calculated (for $M\!\!\ge$0.4\,M$_{\odot}$).
Colours are as in Fig.\,\ref{taugxage} and triangles ($\blacktriangle$, $\triangle$)
are explained in the main text, Section~\ref{parametrizations}.}
\label{roxrzams}
}
\end{figure}

In order to set an alternative location to calculate the local convective turnover
time for fully convective stars during the MS and pre-MS phases, we investigate the
stellar interior properties related to this quantity, evaluating the Rossby number,
the convective velocity and the local convective turnover time itself throughout the
whole star. First, we analyse how Rossby numbers vary inside the stars for all models
at the ZAMS.  From Fig.\,\ref{roxrzams}, we can see that the Rossby number profiles
seem to be very steep, when plotted on a length scale covering several times the
extensions of their corresponding convective zones. Actually, these profiles have
steep inclinations only near the stellar centre and the stellar surface, being
considerably less steep at intermediate stellar radii.  Here, ${\rm Ro}$ was obtained
with $\tau_{\rm c}$ and $P_{\rm rot}$ of each radial shell of the star.  Because our
models include differential rotation, $P_{\rm rot}$ varies with the distance relative
to the star's centre.  In this plot, we identify with crosses the standard location
inside the stars where $\tau_{\rm c}$ is usually calculated. We note that, for each
model, this coincides with the less steep region of the ${\rm Ro}$ radial profile.

\begin{figure} %Fig 5
\centering{
\includegraphics[width=\columnwidth]{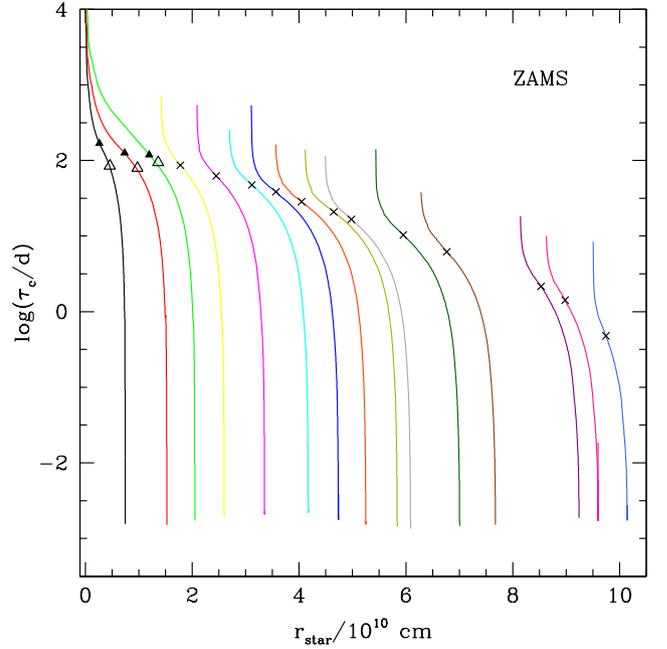}
\caption{Local convective turnover time $\tau_{\mathrm{c}}$ {\it versus} radius
$r_{\mathrm{star}}$ for each stellar model.  Symbols are as in Fig.\,\ref{roxrzams}
and colours are as in Fig.\,\ref{taugxage}.}
\label{taulxrzams}
}
\end{figure}

The behaviour of ${\rm Ro}$ with radius at the ZAMS, shown in Fig.\,\ref{roxrzams},
depends on the convective turnover time profiles (Fig.\,\ref{taulxrzams}) which, in
turn, depend on convective velocity profiles, shown in Fig.\,\ref{vcxrzams}. Both
curves also vary steeply when plotted on large length scales and $r_{\rm std}$
localises around the less steep regions of the profiles, similar to ${\rm Ro}$.

\subsection{Parameterisations of $r_{\rm std}$ at the ZAMS}\label{parametrizations}

\begin{figure} %Fig 7
\centering{
\includegraphics[width=\columnwidth]{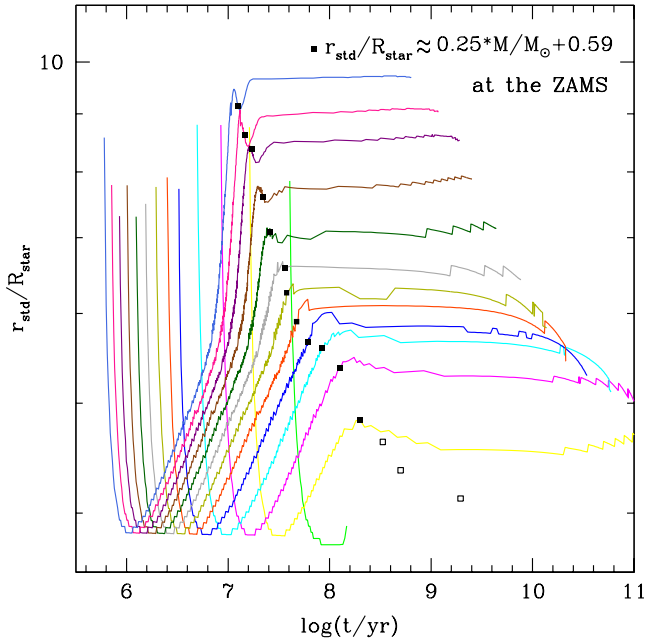}
\caption{%
The standard location where $\tau_{\rm c}$ is calculated in units of stellar radius
{\it versus} age $t$ for each stellar model. Filled squares ($\blacksquare$)
represent these locations at the ZAMS for $M\!\!\ge$0.4\,M$_{\odot}$. Open squares
($\square$) represent extrapolations to 0.1, 0.2 and 0.3\,M$_{\odot}$.  Colours are
as in Fig.\,\ref{taugxage}.}
\label{distraioz}
}
\end{figure}

Continuing our effort to find an alternative place to calculate $\tau_{\rm c}$ inside
fully convective stars, we next express the standard location in terms of other
quantities, such as the stellar radius $R_{\rm star}$ and the pressure scale height
$H_{\rm p}$.  In Fig.\,\ref{distraioz}, we show these standard locations as a
function of age and mass for models with $M\!\!\ge$0.4\,M$_{\odot}$.

The standard location $r_{\rm std}$ varies from about 0.65 to 0.95\,$R_{\rm star}$ in
the mass range of 0.4
\begin{figure} %Fig 6
\centering{
\includegraphics[width=\columnwidth]{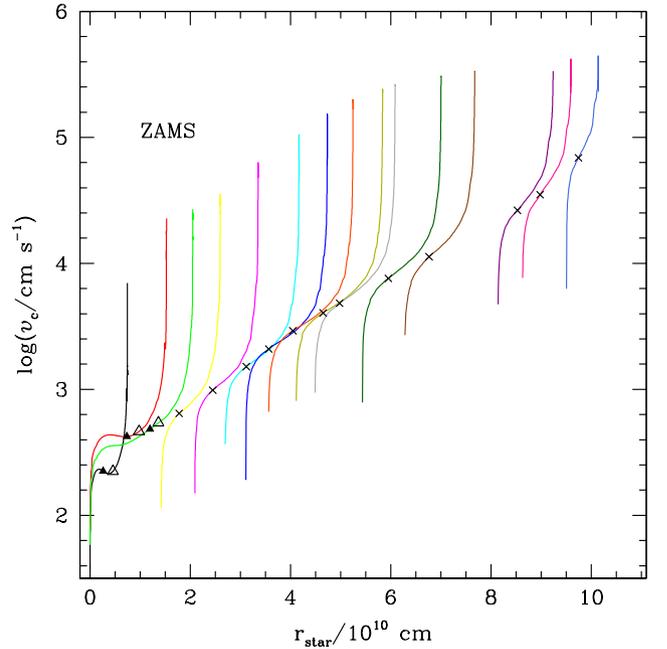}
\caption{Convective velocity $\varv_{\mathrm{c}}$ {\it versus} the stellar radius
$r_{\mathrm{star}}$ for each mass model. Symbols are as in Fig.\,\ref{roxrzams} and
colours as in Fig.\,\ref{taugxage}.}
\label{vcxrzams}
}
\end{figure}
to 1.5\,M$_{\odot}$ at the ZAMS.  Black squares ($\blacksquare$) indicate $r_{\rm
std}$ inside the stars at the ZAMS.  We analysed how $r_{\rm std}/R_{\rm star}$
varies as a function the stellar mass and found that it behaves approximately
linearly. We then made a linear fit to the data obtaining $r_{\mathrm{std}}/R_{\rm
star}=(0.25\pm 0.01)M/\mathrm{M}_{\odot}+(0.59 \pm 0.01)$.  Through an extrapolation
of this fit to the lower mass regime, we estimate that the local convective turnover
time can be calculated as 0.615, 0.640 and 0.665\,R$_{\rm star}$ respectively for
0.1, 0.2 and 0.3\,M$_{\odot}$ above the base of the convective zone at the ZAMS.
These extrapolations are shown in Fig.\,\ref{distraioz} as open squares ($\square$).
Aiming to visualise where our predicted positions to calculate $\tau_{\rm c}$ at the
ZAMS for 0.1, 0.2 and 0.3\,M$_{\odot}$ would be located relative to the ${\rm Ro}$,
$\tau_{\rm c}$ and $\varv_{\rm c}$ profiles, we marked them, respectively in
Figs\,\ref{roxrzams}, \ref{taulxrzams} and \ref{vcxrzams} with black open triangles
($\triangle$).  The trends followed by our predicted positions in the ${\rm Ro}$ and
$\varv_{\rm c}$ profiles are similar to those corresponding to the locations of
$r_{\rm std}$ for $M\!\!\ge\!\!0.4$\,M$_{\odot}$ (crosses - $\times$). However, for
the $\tau_{\rm c}$ profile this behaviour is not seen and results in predicted
positions that are below expectations and that do not change significantly with mass
in the range of 0.1$\le M/M_{\odot} \le$0.3.

\begin{figure} %Fig 8
\centering{
\includegraphics[width=\columnwidth]{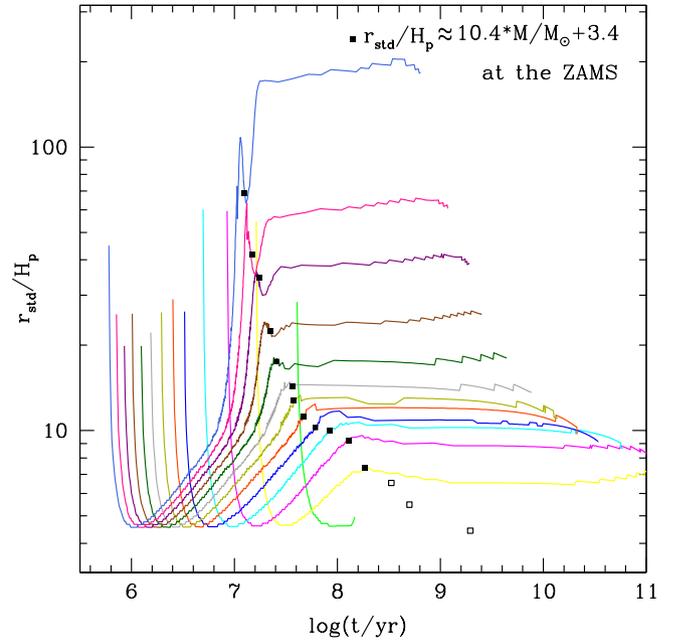}
\caption{%
The standard location where $\tau_{\rm c}$ is calculated in units of $H_{\rm p}$ {\it
versus} the stellar age $t$ for some models. Symbols are as in Fig.\,\ref{distraioz}
and colours are as in Fig.\,\ref{taugxage}.
}
\label{disthpz}
}
\end{figure}

Repeating this procedure for $H_{\rm p}$ we show, in Fig.\,\ref{disthpz}, the
standard locations where $\tau_{\rm c}$ should be calculated in terms of $H_{\rm p}$
as a function of age and mass for models with $M\!\!\ge$0.4\,M$_{\odot}$. Black
squares ($\blacksquare$) indicate $r_{\rm std}$ inside the stars at the ZAMS.  We
note that $r_{\rm std}$ lies in the interval of 7 to 70\,H$_{\rm p}$ in the mass
range of 0.4 to 1.5\,M$_{\odot}$ at the ZAMS. The variation of $r_{\rm std}/H_{\rm
p}$ with mass deviates more from a linear behaviour than does $r_{\rm std}/R_{\rm
star}$, mainly for $M\!\!>\!\!1.0$\,M$_{\odot}$. For this reason, we performed a
linear fit with $r_{\rm std}/H_{\rm p}$ and $M$ for $M\!\!\le$1.0\,M$_{\odot}$
finding $r_{\rm std}/R_{\mathrm{star}}= (10.4 \pm 0.9)M/\mathrm{M_{\odot}}+(3.4 \pm
0.7)$.  An extrapolation of this fit to lower masses indicates that the local
convective turnover time should be calculated at 4.44, 5.48 and 6.53\,H$_{\rm p}$,
respectively for 0.1, 0.2 and 0.3\,M$_{\odot}$ above the base of the convective zone
at the ZAMS. These extrapolations are shown in Fig.\,\ref{disthpz} as open squares
($\square$) and their corresponding locations relative to the ${\rm Ro}$, $\tau_{\rm
c}$ and $\varv_{\rm c}$ profiles in Figs\,\ref{roxrzams}, \ref{taulxrzams} and
\ref{vcxrzams} respectively are marked with black filled triangles
($\blacktriangle$). In these latter Figures, we can see that our predicted positions
in the $\rm Ro$, $\tau_{\rm c}$ and $\varv_{\rm c}$ profiles follow the same trends
as the corresponding locations of $r_{\rm std}$ values for
$M\!\!\ge\!\!0.4$\,M$_{\odot}$ (crosses - $\times$). Our extrapolated positions to
calculate $\tau_{\rm c}$ were compared to those predicted by \citet{feiden13}.  He
used results of full 3D MHD models of fully convective stars (0.3\,M$_{\odot}$)
obtained by \citet{browning08}. These suggest that the magnetic field strength is
maximum at about 0.15$R_{\rm star}$, and \citet{feiden13} assumed that this radial
position plays the same role as the tachocline in partly convective stars. So, he
calculated the Rossby number at one half of a mixing length above 0.15$R_{\rm star}$.
In order to make this comparison, we used the values of $R_{\rm star}$ and $\ell$
yielded by the {\ttfamily ATON} code for each stellar mass.  Both locations are, to a
certain degree, consistent with each other, because they are found in the less steep
region of the ${\rm Ro}$, $\tau_{\rm c}$ and $\varv_{\rm c}$ profiles; our predicted
positions are about 10\% smaller than those of \citet{feiden13} but the places where
our tachocline-like region should be are 0.37, 0.45 and 0.51\,R$_{\rm star}$,
respectively for 0.1, 0.2 and 0.3\,M$_{\odot}$.

We next investigate the stellar interior near the point where $\tau_{\rm c}$ should
be calculated (both standard and predicted positions) aiming to find some
correlations with the stellar properties.  In Fig.\,\ref{taulxrzams}, we note that
the standard positions (defined for $M$$\ge$0.4M$_{\odot}$) and our predicted
positions (defined for $M$$<$0.4M$_{\odot}$) are found near the inflection point of
the curves, as was confirmed by 2nd derivative calculations of 3rd degree spline
fits.  In order to have a better comprehension of this coincidence it would be useful
to inspect results of MHD simulations, which can provide us important informations
about the dynamo process in such regions. 

As can be seen in Fig.\,\ref{distraioz}, $r_{\rm std}/R_{\rm star}$ does not vary
significantly after the ZAMS.  From Fig.\,\ref{disthpz}, the same behaviour is found
for $r_{\rm std}/H_{\rm p}$ for $M\!\! \le$1.3\,M$_{\odot}$ but for larger masses the
roughly constant MS $r_{\rm std}/H_{\rm p}$ are higher than the corresponding ZAMS.
At first, this could indicate that the $r_{\rm std}/R_{\rm star}$ scaling should be
favoured over the $r_{\rm std}/H_{\rm p}$ scaling to estimate the location where
$\tau_{\rm c}$ should be calculated but on the other hand, it underestimates such
positions for stars in the most critical mass regime of our analysis
($M$$\leq$0.3\,M$_{\odot}$, in which stars are fully convective throughout their
entire evolution). Taking this fact into account, hereinafter we will adopt the
description of $r_{\rm std}$ in terms of $H_{\rm p}$.

\begin{figure} %Fig 9
\centering{
\includegraphics[width=\columnwidth]{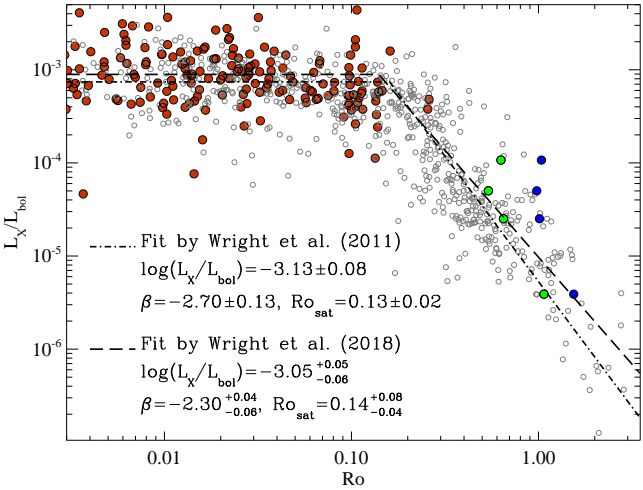}
\caption{Fractional X-ray luminosity as a function of the Rossby number for 828 stars
in the samples of \citet{wright11} and \citet{wright16}. Fast rotating fully
convective stars are shown as filled dark red circles
(\textcolor{dred}{$\bullet$}\hspace{-0.155cm}$\circ$) and partially convective stars
as grey open circles (\textcolor{grey}{$\circ$}).  The four slowly rotating fully
convective stars are shown with ${\rm Ro}$ calculated with two different $\tau_{\rm
c}$ values, semi-empirically (filled green circles,
\textcolor{green}{$\bullet$}\hspace{-0.155cm}$\circ$) or theoretically according to
this work (filled blue circles, \textcolor{blue}{$\bullet$}\hspace{-0.150cm}$\circ$).
The dash-dotted and long-dashed lines show the best-fitting activity-rotation
relationships found by \citet{wright11} and \citet{wright18}, respectively.}
\label{figwright16}
}
\end{figure}

Then, we used our $r_{\rm std}/H_{\rm p}$ linear fit as a function of mass at the
ZAMS to estimate $\tau_{\rm c}$ for the 4 slowly rotating fully convective stars in
the sample of \citet{wright16}, the first to be found in the unsaturated region of
the activity-rotation diagram.  Fig.\,\ref{figwright16} illustrates this, showing
$L_{\rm X}/L_{\rm bol}$\,{\it versus}\,${\rm Ro}$ for 828 stars.  Among them 824 are
from the sample of \citet{wright11}, which is composed by partially convective stars
(\textcolor{grey}{$\circ$}) and fast rotating fully convective stars
(\textcolor{dred}{$\bullet$}~\hspace{-0.223cm}$\circ$). Fig.\,\ref{figwright16} also
shows the 4 slowly rotating fully convective stars
(\textcolor{green}{$\bullet$}\hspace{-0.165cm}$\circ$) from the sample of
\citet{wright16}.  The dash-dotted line is the best fit found by \citet{wright11} for
the 824 stars in their sample and the long dashed line is the best fit found by
\citet{wright18} for these 824 stars plus 23 slow rotating fully convective stars.
\citet{wright11} and \citet{wright16} used the semi-empirical determinations of local
convective turnover times of \citet{pizzolato03} to calculate the Rossby number of
their sample of stars.  In order to test our preliminary results, we calculated the
Rossby numbers of these 4 fully convective stars, in the sample of \citet{wright16},
using our theoretical $\tau_{\rm c}$. Their positions in the rotation-activity
diagram obtained with our $\tau_{\rm c}$ are shown in blue filled circles
(\textcolor{blue}{$\bullet$}\hspace{-0.170cm}$\circ$) in Fig.\,\ref{figwright16}.
Though our ${\rm Ro}$ are slightly higher about 1.5 times) than those of
\citet{wright16}, they still keep the corresponding stars in the unsaturated region.

\subsection{Parameterisations of $r_{\rm std}$ at different ages}\label{parametages}

After achieving promising results with the previous fit and $\tau_{\rm c}$ estimate,
we repeated the procedure described in Section~\ref{parametrizations} for some
stellar ages $t$, namely, $\log(t/{\rm yr})\in\{$6.0, 6.5, 7.0, 7.5, 8.0, 8.5, 9.0,
9.5, 10.0, 10.14$\}$.  For each of these ages, we performed a linear fit to $r_{\rm
std}/H_{\rm p}$ as a function of the stellar mass.  Table\,\ref{tabfits} shows
details of each linear fit for a given age.  For $\log(t/{\rm yr})\!\!  \le \!\!6.0$,
we used the fit obtained for $\log(t/{\rm yr}) \!\!=\!\!6.0$ to find the location to
calculate $\tau_{\rm c}$. For ages in the interval of $6.0\!\!<\!\!  \log(t/{\rm
yr})\!\!  \le \!\!6.5$, we interpolate $r_{\rm std}/H_{\rm p}$ with the fits obtained
for $\log(t/\rm yr) \!\!=\!\!6.0$ and $\log(t/\rm yr) \!\!=\!\!6.5$. For ages in the
interval of $6.5\!\! < \!\!  \log(t/{\rm yr}) \!\! \le \!\! 7.0$, we made a similar
interpolation but with fits found at $\log(t/{\rm yr}) \!\!=\!\!6.5$ and $\log(t/{\rm
yr}) \!\!=\!\!7.0$. The next intervals were divided in the same way, every 0.5~dex,
except the last one, for which the interpolation was done with the fits obtained at
$\log(t/{\rm yr}) \!=\!10.0$ and $\log(t/{\rm yr}) \!=\!10.14$. The mass interval
used in our linear fits differs for each age. The minimum mass in each mass range is
0.4\,M$_{\odot}$, except in cases in which the models have large mixing lengths
placing $r_{\rm std}$ outside the stars. This happens for $\log(t/{\rm yr}) \!\! =
\!\!6.0$ below 1.3\,M$_{\odot}$, $\log(t/{\rm yr}) \!\! = \!\!6.5$ below
0.9\,M$_{\odot}$ and $\log(t/{\rm yr}) \!\! = \!\!7.0$ below 0.6\,M$_{\odot}$.
Because the relations between $r_{\rm std}/H_{\rm p}$ and $M$ deviate more from a
linear behaviour when more massive stars are included, some stellar masses were
excluded from the fits. The exceptions are $\log(t/{\rm yr}) \!\! = \!\!10.0$, and
$\log(t/{\rm yr}) \!\! = \!\!10.14$, for which stars with masses larger\, than\,
1.0~M$_{\odot}$ and\, 0.9~M$_{\odot}$,\, respectively,\, have\, already left the MS
phase.

\begin{table}
\caption{Details of linear fits with the equation ${r_{\rm std}}/{H_{\rm p}}=A\,
M/M_{\odot}+B$.  Column\,1 gives the logarithm of stellar age; col.\,2 the mass range
used in the fit of $r_{\rm std}/H_{\rm p}$; col.\,3 the coefficient A; and col.\,4
the coefficient B.} 
\label{tabfits}
\centering
{\footnotesize
\advance\tabcolsep by -3pt
\begin{tabular}{rccc}
\hline \hline
log($t/{\rm yr}$)         &  Mass range      & coefficient      & coefficient \\ 
                 & M$_{\odot}$    & A & B \\ \hline
\hline
 6.00  & 1.3 to 1.5  &  \hphantom{1.}$-6\pm2$\hphantom{1.}    & \hphantom{.}$14\pm3$\hphantom{1.}    \\ [-1.5pt]
 6.50  & 0.9 to 1.5  &  \hphantom{1}$3.6\pm0.3$  & $1.1\pm0.3$ \\ [-1.5pt]
 7.00  & 0.6 to 1.2  &  \hphantom{1}$7.2\pm0.3$  & $0.0\pm0.3$ \\ [-1.5pt]
 7.50  & 0.4 to 0.9  &  \hphantom{0.}$13\pm1$\hphantom{1.}     & $7.2\pm0.8$ \\ [-1.5pt]
 8.00  & 0.4 to 1.2  &  \hphantom{0.}$18\pm3$\hphantom{1.}     & \hphantom{}$-1\pm2$\hphantom{1.}   \\ [-1.5pt]
 8.50  & 0.4 to 1.2  &  \hphantom{0.}$17\pm3$\hphantom{1.}     & \hphantom{}$-1\pm2$\hphantom{1.}   \\ [-1.5pt]
 9.00  & 0.4 to 1.1  &  \hphantom{0.}$14\pm1$\hphantom{1.}     & \hphantom{0.}$1\pm1$\hphantom{1.}    \\ [-1.5pt]
 9.50  & 0.4 to 1.1  &  \hphantom{0.}$13\pm2$\hphantom{1.}     & \hphantom{0.}$1\pm1$\hphantom{1.}     \\ [-1.5pt]
10.00  & 0.4 to 1.0  &  \hphantom{10.}$8\pm2$\hphantom{1.}      & \hphantom{0.}$4\pm1$\hphantom{1.}     \\ [-1.5pt]
10.14  & 0.4 to 0.9  &  \hphantom{10.}$9\pm2$\hphantom{1.}      & \hphantom{0.}$4\pm1$\hphantom{1.}     \\ [-1.5pt]
ZAMS   & 0.4 to 1.0  &  $10.4\pm0.9$ & $3.4\pm0.7$ \\ [-1.5pt] \hline
\end{tabular}
}
\end{table}

We analysed the radial profiles of $\tau_{\rm c}$ and $\varv_{\rm c}$ for all
selected ages in Table~\ref{tabfits}.  Their behaviours are similar to those seen for
these quantities at the ZAMS (Figs.\,\ref{taulxrzams} and \ref{vcxrzams}). The
predicted positions where $\tau_{\rm c}$ should be calculated for models with
$M\!\!<\!\!0.4$M$_{\odot}$, which are outside the interpolations intervals shown in
Table~\ref{tabfits}, follow the same trend as the corresponding positions of $r_{\rm
std}$. They are found to be near the place where the curves are less steep. Because
the fit for $\log(t/{\rm yr}) $$=$6.0 was made with the smallest number of points, it
is the one with the largest departure from this trend. The radial $\tau_{\rm{c}}$
profiles for $\log(t/{\rm yr}) \!\!=\!\!7.0$ present the same behaviour found as at
the ZAMS, that the standard positions lie very close to the inflection point of the
curves, as do our predicted positions. This was confirmed by calculations with 3rd
degree spline fits. The analysis of the radial $\tau_{\rm c}$ profiles is postponed
to future work, when we will try to interpret our findings in the light of MHD
simulation results.

\section{Results} \label{results}

The results obtained in Section\,3 through linear fits of $r_{\rm std}/H_{\rm p}$ as
a function of mass were introduced in the {\ttfamily ATON} code as alternative
locations to calculate $\tau_{\rm c}$ for stars where
$r_\mathrm{std}$$>$$R_\mathrm{star}$, for fully convective stars. Then, the evolution
of local convective turnover times during the pre-MS and at the beginning of the MS
was followed for 0.1 to 1.5\,M$_{\odot}$ stars and tabulated together with the
corresponding evolutionary tracks. Table~\ref{tabtrack} presents the 1\,M$_{\odot}$
model as an example of such tables.  Rotation periods are those evaluated by models
for the surface layers, starting from the initial periods, which correspond to the
initial angular momenta given in Eq.\,1, and then evolving by considering
conservation of angular momentum. In our models, we did not take into account angular
momentum loss by stellar winds. This is the main reason why our 1\,M$_{\odot}$ model
at the solar age shows a rotation period quite different from the current Sun. In
order to reproduce it, we should have used the rotational scheme 3, described in
Section~\ref{models}, but we chose scheme 2 because differential rotation and its
interaction with convection is the base of the solar dynamo model.  To have an idea
of how the choice of rotational scheme affects convective turnover time, we compared
$\tau_{\rm c}$ for 1\,M$_{\odot}$ at the Sun's age generated by models considering
differential rotation throughout the whole star (scheme 2) and models considering
differential rotation in the radiative core and rigid body rotation in the convective
envelope (scheme 3) plus angular momentum loss by stellar winds.  The difference
between them was 5\%.

\begin{table}
\caption{Evolutionary tracks, including $\tau_{\rm c}$, $\tau_{\rm g}$, $P_{\rm rot}$
and ${\rm Ro}$, for a 1\,M$_{\odot}$ star$^a$.  Column\,1 gives the logarithm of
stellar age; col.\,2 the logarithm of bolometric luminosity; col.\,3 the logarithm of
effective temperature; col.\,4 the logarithm of effective gravity; col.\,5 the
logarithm of local convective turnover times; col.\,6 the logarithm of global convective turnover
times; col.\,7 the rotation period; and col.\,8 the Rossby number.
}
\label{tabtrack}
\centering
{\small
\advance\tabcolsep by -4.0pt
\begin{tabular}{rrrcrrrc}
\hline \hline
${{\log(t/{\rm yr})}}\vphantom{\Big|}$ & $\log\frac{L}{L_{\odot}}$ &
${\log}\atop{(T_{\rm eff}/{\rm K)}}$ & ${\log}\atop{(g/{\mathrm{cm\,s^{-2}}})}$ &
${{\log}\atop{(\tau_{\rm c}/{\rm d)}}}$ &
${{\log}\atop{(\tau_{\rm g}/{\rm d})}}$ & ${P_{\rm rot}/{\rm d}}$ & Ro \\ \hline
	&   &  &  &  &  &  & \\ [-8pt]
 2.6848 &    1.6882 & 3.6059 & 2.126 & 1.9444 & 2.8296 & 185.776 & 2.1113 \\ [-1.5pt]
 3.8668 &    1.5802 & 3.6337 & 2.345 & 1.9110 & 2.7955 & 103.295 & 1.2680 \\ [-1.5pt]
 4.3506 &    1.3600 & 3.6441 & 2.607 & 1.9006 & 2.7781 &  58.140 & 0.7309 \\ [-1.5pt]
 4.8159 &    1.1448 & 3.6526 & 2.856 & 1.8853 & 2.6430 &  33.698 & 0.4388 \\ [-1.5pt]
 5.2182 &    0.9844 & 3.6574 & 3.036 & 1.8858 & 2.7335 &  22.858 & 0.2973 \\ [-1.5pt]
 5.3960 &    0.8181 & 3.6611 & 3.217 & 1.8821 & 2.7695 &  14.976 & 0.1965 \\ [-1.5pt]
 5.6561 &    0.5970 & 3.6633 & 3.447 & 1.8819 & 2.7758 &   8.929 & 0.1172 \\ [-1.5pt]
 5.9360 &    0.3761 & 3.6617 & 3.661 & 1.8865 & 2.8091 &   5.498 & 0.0714 \\ [-1.5pt]
 6.2263 &    0.1553 & 3.6572 & 3.864 & 1.8909 & 2.7400 &   3.477 & 0.0447 \\ [-1.5pt]
 6.5282 & $-$0.0652 & 3.6506 & 4.058 & 2.1183 & 2.5221 &   2.129 & 0.0162 \\ [-1.5pt]
 6.8479 & $-$0.2581 & 3.6476 & 4.239 & 2.0075 & 2.3410 &   1.405 & 0.0138 \\ [-1.5pt]
 7.1062 & $-$0.2979 & 3.6659 & 4.352 & 1.8314 & 2.1384 &   1.037 & 0.0153 \\ [-1.5pt]
 7.2812 & $-$0.1742 & 3.7077 & 4.396 & 1.6152 & 1.9158 &   0.880 & 0.0214 \\ [-1.5pt]
 7.3982 & $-$0.0095 & 3.7492 & 4.397 & 1.3542 & 1.6664 &   0.845 & 0.0374 \\ [-1.5pt]
 7.5163 & $-$0.0708 & 3.7640 & 4.517 & 1.2124 & 1.5173 &   0.648 & 0.0397 \\ [-1.5pt]
 9.0239 & $-$0.0894 & 3.7605 & 4.522 & 1.2321 & 1.5315 &   0.659 & 0.0386 \\ [-1.5pt]
 9.4180 & $-$0.0408 & 3.7644 & 4.489 & 1.2148 & 1.5253 &   0.704 & 0.0429 \\ [-1.5pt]
 9.5787 &    0.0011 & 3.7674 & 4.459 & 1.1977 & 1.5018 &   0.747 & 0.0474 \\ [-1.5pt]
 9.6992 &    0.0445 & 3.7699 & 4.425 & 1.1819 & 1.4831 &   0.804 & 0.0529 \\ [-1.5pt]
 9.8050 &    0.1019 & 3.7723 & 4.378 & 1.1707 & 1.4858 &   0.888 & 0.0599 \\ \hline
\multicolumn{8}{p{0.95\columnwidth}}{$^a$The complete version of the table, including
15 tracks for masses in the range 0.1-1.5\,M$_{\odot}$ (in 0.1\,M$_{\odot}$ 
increments), will be available in electronic form at CDS via anonymous ftp to 
cdsarc.u-strasbg.fr (130.79.128.5) or via https://cdsarc.unistra.fr/viz-bin/cat/J/MNRAS.
Remind that in this work all the logarithms are taken in base 10.}
\end{tabular}
}
\end{table}

According to our 1\,M$_{\odot}$ model at the age of the Sun, the solar local
convective turnover time is $\tau_{\rm c,\odot}$$=$15.40~d which corresponds to
${\rm Ro_{\odot}}$$=$1.69 with $P_{\rm rot,\odot}$$=$26.09~d \citep{donahue96}. This
value is consistent to that found semi-empirically by \citet{pizzolato03}, which is
$\tau_{\rm c,\odot}$$=$12.59~d (${\rm Ro_{\odot}}$$=$2.07). 

\begin{figure} %Fig 10
\centering{
\includegraphics[width=\columnwidth]{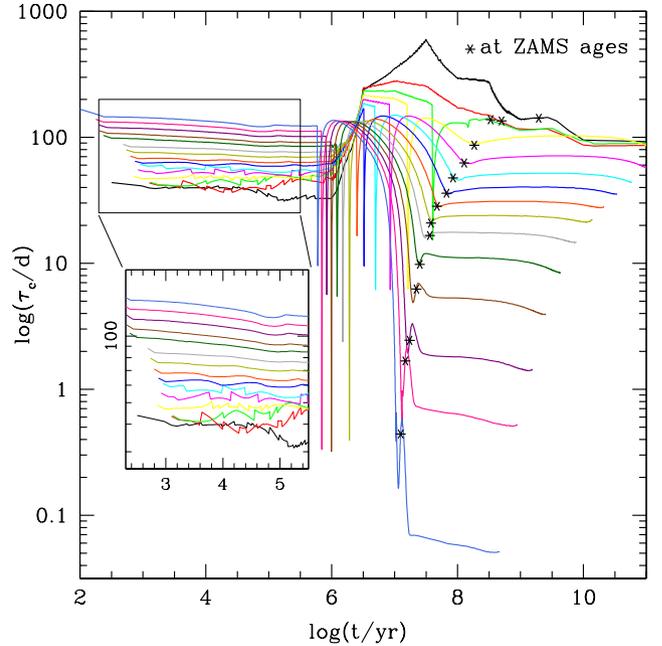}
\caption{As Fig.\,\ref{taucxage} but extended to include $\tau_{\rm c}$ obtained with
our alternative location described in Section~\ref{calculations} for models in which
$r_{\rm std}$$>$$R_{\rm star}$. The insert shows in detail the temporal evolution of
$\tau_{\rm c}$ during the beginning of the pre-MS phase.
}
\label{taucxagenewloc}
}
\end{figure}

Fig.\,\ref{taucxagenewloc}, similarly to Fig.\,\ref{taucxage}, shows local convective
turnover times as a function of age and mass. The ZAMS age of each mass model is
shown with an asterisk.  Compared to Fig.\,\ref{taucxage}, Fig.\,\ref{taucxagenewloc}
is extended, including $\tau_{\rm c}$ calculations where $r_{\rm std}\!\!>\!\!R_{\rm
star}$, for $M\!\!<\!\!0.4$\,M$_{\odot}$ (pre-MS and MS) and for the early pre-MS of
stars with $M\!\!\ge\!\!0.4$\,M$_{\odot}$. With regard to the last case, the age at
which $r_{\rm std}$ becomes smaller than $R_{\rm star}$ decreases as the mass
increases, as expected from the stellar interior theory. For the 0.4M$_{\odot}$ model
it is about $\log(t/{\rm yr}) \!\!=\!\!7.0$, while for the 1.5M$_{\odot}$ model it is
about $\log(t/{\rm yr}) \!\!=\!\!6.0$. For models whose $r_{\rm std}\!\!<\!\!R_{\rm
star}$, $\tau_{\rm c}$ was calculated at one-half of a mixing length above the base
of the convective zone.  Otherwise $\tau_{\rm c}$ was calculated at the alternative
place, related to $H_{\rm p}$.
As explained in Section~\ref{calculations}, for $\log(t/{\rm yr}) \!\!\le\!\!6.0$ we
use the same $r_{\rm std}/H_{\rm p}$ linear fit found at $\log(t/{\rm yr})
\!\!=\!\!6.0$. This is why $\tau_{\rm c}$ is approximately constant in this age
range. For $\log(t/{\rm yr}) \!\!\le\!\!6.0$ and $M\!\!\ge\!\!0.4$\,M$_{\odot}$, the
larger the stellar mass the larger $\tau_{\rm c}$, while for
$M\!\!<\!\!0.4$\,M$_{\odot}$ we cannot observe a regular behaviour (see the amplified
view in Fig.\,\ref{taucxagenewloc}). For $\log(t/{\rm yr}) \!\!\ge\!\!7.0$,
$\tau_{\rm c}$ increases as the stellar mass decreases, except for the mass range 0.1
to 0.4\,M$_{\odot}$ and $\log(t/{\rm yr}) \gtrsim 9.3$. We believe that this
irregular behaviour of models with $M$$<$0.4\,M$_{\odot}$ is probably a consequence
of the method we used to determine where $\tau_{\rm c}$ should be calculated. This
was based on linear fits, for different age intervals, excluding higher masses
because they deviate more from the linear behaviour.

\begin{figure} %Fig 11
\centering{
\includegraphics[width=\columnwidth]{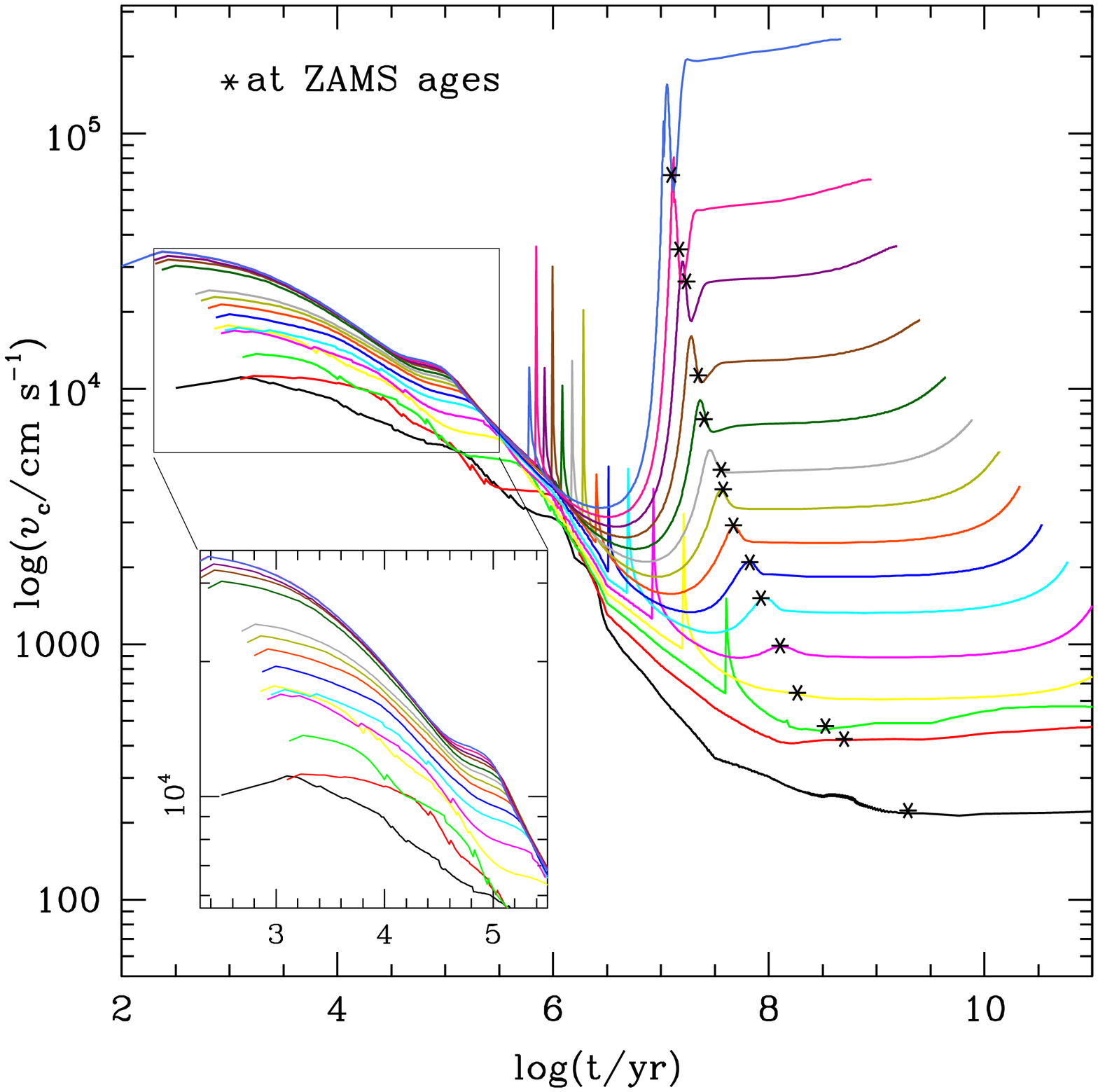}
\caption{Same as in Fig.\,\ref{velcxage} but extended to include $\varv_{\rm c}$
obtained with our alternative location described in Section~\ref{calculations} for
models in which $r_{\rm std}$$>$$R_{\rm star}$. The insert shows in detail the
temporal evolution of $\varv_{\rm c}$ during the beginning of the pre-MS phase.}
\label{velcxagenewloc}
}
\end{figure}

Fig.\,\ref{velcxagenewloc}, which is similar to Fig.\,\ref{velcxage}, shows the
convective velocities used to calculate the $\tau_{\rm c}$ shown in
Fig.\,\ref{taucxagenewloc} and the ZAMS age of each mass model.  In comparison with
Fig.\,\ref{velcxage}, Fig.\,\ref{velcxagenewloc} is extended to include $\varv_{\rm
c}$ for models in which $r_{\rm std}\!\!>\!\!R_{\rm star}$.  On the MS, the larger
the stellar mass the larger the convective velocity.  The same behaviour is found on
the pre-MS, except for models with $M$$<$0.3\,M$_{\odot}$ and in the mass range 0.4
to 0.6\,$M_{\odot}$ for ages smaller than $10^4$\,yr (see the amplified view of
Fig.\,\ref{velcxagenewloc}). In the age range 5.5$\leq$$\log(t/{\rm yr})$$\leq$8.0,
where radiative cores form, this general correlation between $v_{\rm c}$ and age is
modified essentially by the peaks in the $v_{\rm c}$ curves.

As $\tau_{\rm c}$, through its relation with $\rm Ro$, plays an important role in
stellar activity investigations and is closely related to the stellar mass, we
parameterise our updated local convective turnover times by fitting linearly
$\log(\tau_{\rm c})$ as a function of mass for each age shown in Table~\ref{tabfits}.
The best fit for $\log(\tau_{\rm c}/{\rm d})$ at the ZAMS is
\begin{equation}
\log(\tau_{\rm c}/{\rm d})=(-1.19\pm 0.02)\, M/M_{\odot}+(2.40\pm 0.02).
\label{eqtaufitz}
\end{equation}

\noindent
This is valid over the range 0.4\,M$_{\odot}$$\le$$M$$\le$1.0\,M$_{\odot}$. Similar
parameterisations were made for the other ages shown in Table~\ref{tabfits}.
Table~\ref{tabfitstau} shows details of all these linear fits. 

\begin{table}
\caption{Details of linear fits with the equation $\tau_{\rm c}=A\, M/M_{\odot}+B$.
Column\,1 gives the logarithm of stellar age, col.\,2 the mass range used in the fits
(the same used in Table\,\ref{tabfits}), col.\,3 the coefficient A and col.\,4 the
coefficient B.}
\label{tabfitstau}
\centering
{\footnotesize
\advance\tabcolsep by -3pt
\begin{tabular}{rccc}
\hline \hline
log($t/{\rm yr}$)         &  Mass range      & coefficient      & coefficient \\ 
                 & M$_{\odot}$    & A & B \\ \hline
\hline
 6.00  & 1.3 to 1.5  &  \hphantom{11}$0.6\pm 0.2$\hphantom{.}    & $1.3\pm0.2$   \\ [-1.5pt]
 6.50  & 0.9 to 1.5  &  $-0.30\pm0.03$  & $2.41\pm0.03$ \\ [-1.5pt]
 7.00  & 0.6 to 1.2  &  $-0.78\pm0.05$  & $2.67\pm0.04$ \\ [-1.5pt]
 7.50  & 0.4 to 0.9  &  $-1.4\pm0.2$    & $2.8\pm0.1$   \\ [-1.5pt]
 8.00  & 0.4 to 1.2  &  $-1.5\pm0.1$    & $2.58\pm0.08$ \\ [-1.5pt]
 8.50  & 0.4 to 1.2  &  $-1.4\pm0.1$    & $2.58\pm0.09$ \\ [-1.5pt]
 9.00  & 0.4 to 1.1  &  $-1.36\pm0.05$  & $2.54\pm0.04$ \\ [-1.5pt]
 9.50  & 0.4 to 1.1  &  $-1.41\pm0.07$  & $2.57\pm0.05$ \\ [-1.5pt]
10.00  & 0.4 to 1.0  &  $-1.26\pm0.07$  & $2.47\pm0.05$ \\ [-1.5pt]
10.14  & 0.4 to 0.9  &  $-1.36\pm0.05$  & $2.53\pm0.03$ \\ [-1.5pt]
ZAMS   & 0.4 to 1.0  &  $-1.19\pm0.02$  & $2.40\pm0.02$ \\ [-1.5pt] \hline
\end{tabular}
}
\end{table}

\section{Applications and comparisons with observations} \label{applications}

\begin{figure} %Fig 12
\centering{
\includegraphics[width=\columnwidth]{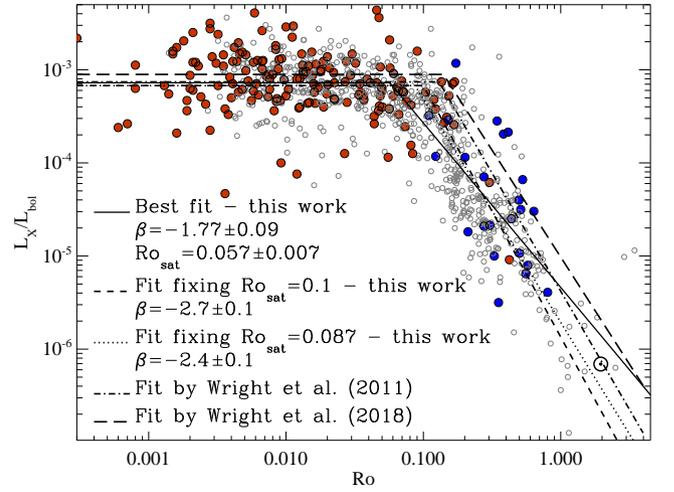}
\caption{Fractional X-ray luminosity as a function of the Rossby number for 847
stars, 824 from \citet{wright11} and 23 slowly rotating fully convective stars from
\citet{wright18}.  Partially convective stars are shown as grey open circles
(\textcolor{grey}{$\circ$}), rapidly rotating fully convective stars are shown as
filled dark red circles (\textcolor{dred}{$\bullet$}\hspace{-0.155cm}$\circ$) and
slowly rotating fully convective stars are shown as filled blue circles
(\textcolor{blue}{$\bullet$}\hspace{-0.155cm}$\circ$).  ${\rm Ro}$ was determined
with our theoretical $\tau_{\rm c}$.  The solid line shows our best fit to the data,
the dashed line shows our fit when setting ${\rm Ro}_{\rm sat}$=0.1, the dotted line
shows our fit when setting ${\rm Ro}_{\rm sat}$=0.087, the dash-dotted line shows the
best fit found by \citet{wright11} and the long dashed line shows the best fit found
by \citet{wright18}.}
\label{lxlbolro}
}
\end{figure}

In order to test our theoretical convective turnover times, we used them to calculate
the Rossby number of 847 stars, 824 from the sample of \citet{wright11} and 23 slowly
rotating fully convective stars from that of \citet{wright18} including those 4 from
\citet{wright16}.  First, we estimated a mass and an age for all stars, using as many
of stellar parameters given by \citet{wright11} as possible and observational data
from \citet{wright18}. In this way, we minimise the impact on $\tau_{\rm c}$, besides
those inherent to the way they were obtained - semi-empirically (in
\citealt{wright11,wright18}) or theoretically by us.  For the sample of
\citet{wright11}, we used their effective temperatures and luminosities, except for
the stars for which they signalled a null luminosity and those whose positions in the
HR diagram fell below our tracks. For these stars, which are mainly very low-mass, we
recalculated their luminosities using the \citet{pecaut13} $V$ band bolometric
corrections in function of $V$$-$$K$. For \citet{wright18}'s sample, we calculated
the effective temperatures in function of $V$$-$$K$ using the colour-temperature
relations of \citet{pecaut13}. Luminosities were obtained using $K$ band bolometric
corrections as a function of $V$$-$$J$ of \citet{mann15,mann16}. In these procedures,
the extinction laws of \citet{rieke85} were used. After estimating stellar masses and
ages, we used our models, including the approach described in
Section~\ref{calculations} for fully convective stars, to obtain the local convective
turnover time for our stars sample. In what follows, we used our theoretical
$\tau_{\rm c}$ and the observed rotation periods \citep[obtained
by][]{wright11,wright18} to calculate the Rossby number of each star.  With our
Rossby numbers and the fractional X-ray luminosities of \citet{wright11,wright18}, we
plotted these data in the $L_{\rm X}/L_{\rm bol}$\,{\it versus}\,${\rm Ro}$ plane,
shown in Fig.\,\ref{lxlbolro}, to investigate the magnetic activity-rotation
relationship. Symbols are the same as in Fig.\,\ref{figwright16}.

We fit the distribution of stars in the rotation-activity diagram
(Fig.\,\ref{lxlbolro}) with the following two-part power-law function:
\begin{equation}
\dfrac{L_{\rm X}}{L_{\rm bol}}  =
\begin{cases}
C\,{\rm Ro^{\beta},}  & {\rm \text{ if } {\rm Ro} > {\rm Ro}_{\rm sat},} \\
\left(\dfrac{L_{\rm X}}{L_{\rm bol}}\right)_{\rm sat}, & {\rm \text{ if } {\rm Ro}\leq {\rm Ro}_{\rm sat},}\\
\end{cases}
\label{eqtwopart}
\end{equation}

\noindent where ${\rm Ro_{sat}}$ is the Rossby number where the saturation occurs,
$(L_{\rm X}/L_{\rm bol})_{\rm sat}$ is the mean saturation level of fractional X-ray
luminosity, $\beta$ the power-law slope for unsaturated stars and $C$ a constant.

Given an initial guess for ${\rm Ro_{sat}}$, we determined $(L_{\rm X}/L_{\rm
bol})_{\rm sat}$, which is the average of $L_{\rm X}/L_{\rm bol}$ for ${\rm
Ro}$$\leq$${\rm Ro_{sat}}$.  For ${\rm Ro}$$>$${\rm Ro_{sat}}$, we fitted the data
with a linear regression in a log-log space, keeping the constant coefficient fixed,
so that the $L_{\rm X}/L_{\rm bol}$ at ${\rm Ro_{sat}}$ is equal to $(L_{\rm
X}/L_{\rm bol})_{\rm sat}$.

Initially, following \citet{vidotto14} and \citet{jackson10}, we fixed ${\rm
Ro_{sat}}$$=$0.1 and the best fitting slope for the unsaturated part of the
relationship was found to be $\beta$=$-$2.7$\pm$0.1, with a Pearson correlation
coefficient $|\rho|$$=$0.77 and the saturation estimated to be at $\log (L_{\rm
X}/L_{\rm bol})_{\rm sat}$$=$$-$3.17$\pm$0.01 (dashed line in Fig.\,\ref{lxlbolro}).
This is consistent with what was found by \citet{wright18} and matches the results by
\citet{wright11}, within the error bars.

Still imposing $L_{\rm X}/L_{\rm bol}$= $(L_{\rm X}/L_{\rm bol})_{\rm sat}$ at ${\rm
Ro_{sat}}$, we used a least squares method to estimate ${\rm Ro_{sat}}$, for which
the standard deviation of the data with respect to Eq.\ (\ref{eqtwopart}) reaches its
minimum value. We find ${\rm Ro_{sat}}=0.057\pm 0.00$9, $\beta=-1.77\pm 0.09$ (with
$|\rho|$$=$0.82) and $\log (L_{\rm X}/L_{\rm bol})_{\rm sat}$$=$$-$3.14$\pm$0.02
(solid line in Fig.\,\ref{lxlbolro}).  We then make another least squares fit of
Eq.~(\ref{eqtwopart}) without imposing a fixed constant coefficient at ${\rm
Ro_{sat}}$. The ${\rm Ro_{sat}}$ which minimises the standard deviation agrees with
the former within the errors, produces the minimum difference between the constant
coefficients at ${\rm Ro_{sat}}$, and $L_{\rm X}/L_{\rm bol}$ at ${\rm Ro_{sat}}$ and
$\beta$ found in both linear regressions agree with each other within the error bars.
Although this $\beta$ and $\log (L_{\rm X}/L_{\rm bol})_{\rm sat}$ is consistent with
those found in the literature, ${\rm Ro_{sat}}$ is considerably smaller.  Neither of
these two models (solid and dashed lines in Fig.\,\ref{lxlbolro}) fit the Sun's
position (indicated with the solar symbol) in the rotation-activity diagram. For
comparisons, we overplot the fits by \citep[][dash-dotted line]{wright11} and
\citep[][long dashed line]{wright18}, both described in Fig.\,\ref{figwright16}.

Next, by imposing $L_{\rm X}/L_{\rm bol}$= $(L_{\rm X}/L_{\rm bol})_{\rm sat}$ at
${\rm Ro_{sat}}$ in Eq.~\ref{eqtwopart}, we made a least squares fit to the data by
fixing ${\rm Ro_{sat}}$ at a value which offers the best fit to the Sun. By using the
minimum and maximum values of the solar rotation period from \citet{donahue96}, we
obtained a saturation Rossby number of ${\rm Ro_{sat}}$$=$0.087$\pm$0.006.  This fit
is the dotted line in Fig.\,\ref{lxlbolro} and provides $\beta$$=$$-$2.4$\pm$0.1
(with $|\rho|$$=$0.79) and $\log (L_{\rm X}/L_{\rm bol})_{\rm
sat}$$=$$-$3.17$\pm$0.01.  This is consistent with what is found in the literature,
with $\beta$ matching that of \citet{wright18} within the error bars. This fit is
more reliable and more well-founded than the two previous fits.  Besides offering a
better prediction to the solar magnetic activity level, it provides a power-law index
in excellent agreement with that obtained by \citet{wright18} who analysed the same
sample of stars.  Furthermore, this ${\rm Ro_{sat}}$ is still smaller than those
found by \citet{wright11,wright18} but greater than we previously obtained with our
best-fitting. It is still unclear what causes saturation in the activity-rotation
relationship, but one possible explanation is that the saturation parameter ${\rm
Ro_{sat}}$ is related to the stellar dynamo efficiency and that the coronal
saturation originates from saturation of the dynamo.  In that case, finding a smaller
${\rm Ro_{sat}}$ can imply that the dynamo process saturates at a higher dynamo
number ($N_{\mathrm{D}}\propto \mathrm{Ro}^{-2}$).

Finally, we separated our sample in two subgroups, partly and fully convective stars,
and fitted the data with the same ${\rm Ro_{sat}}$ used before. The fit parameters
found for both subgroups agree with each other and are consistent with those found
for the whole sample. This indicates that totally and partially convective stars seem
to operate dynamos that cannot be distinguished through the analysis of the their
rotation-activity relationships.  In such a scenario, the tachocline would not be an
essential ingredient to the dynamo and the combined effects of differential rotation
and the Coriolis force would be enough to amplify the magnetic field. Although it is
believed that large-scale magnetic fields would not settle in convective zones,
recent MHD simulations without a tachocline have given rise to stable magnetic
structures \citep{brown10}.

\section{Conclusions} \label{conclusions}

We present a new set of pre-MS and MS evolutionary tracks, including Rossby numbers,
global and local convective turnover times in the mass range of 0.1 to
1.5\,M$_{\odot}$.  Aiming to overcome a problem concerning the location where
$\tau_{\rm c}$ should be theoretically determined by evolutionary models of fully
convective stars (for which the standard location, $r_{\rm std}$$=$$\ell/2$ above the
base of the convective zone, is not appropriate, because then $r_{\rm std}$$>$$R_{\rm
star}$), we performed computations of $\tau_{\rm c}$ for models with
$M$$\geq$0.4M$_{\odot}$ throughout the whole convective zone of the stars for some
selected stellar ages.  Profiles of ${\rm Ro}$, $\tau_{\rm c}$ and $\varv_{\rm c}$
are very steep near the centre and the surface of the star but they are less steep at
intermediate stellar radii.  Then, for each selected age, we identified $r_{\rm std}$
in these models and made a linear fit to these positions (in units of $H_{\rm p}$) as
a function of stellar mass.  We also parameterise $\tau_{\rm c}$ as a function of
mass for the same stellar ages.  When extrapolated to lower masses
($M$$<$0.4\,M$_{\odot}$), our predicted locations of where $\tau_{\rm c}$ should be
calculated follow the same trend exhibited by models with $M$$\geq$0.4\,M$_{\odot}$.
I.e., they are supposed to be found near the less steep region of the respective
profiles. This suggests that our method is a promising approach.

Next, we introduced $r_{\rm std}/H_{\rm p}(M,{\rm age})$, our alternative location to
calculate $\tau_{\rm c}$, in models for which $r_{\rm std}$$>$$R_{\rm star}$ in the
{\ttfamily ATON} code. This allowed estimating $\tau_{\rm c}$ for fully convective
stars.  At the beginning of the pre-MS, $\tau_{\rm c}$ is roughly constant and
increases with the stellar mass.  It presents a large variation before reaching the
MS and then remains nearly constant again, with $\tau_{\rm c}$ increasing with
decreasing mass.  The convective velocity decreases in the beginning of the pre-MS
and the larger the stellar mass the larger the convective velocity. It varies
significantly before reaching the MS and remains almost constant after that.  On the
MS $\varv_{\rm c}$ increases with the stellar mass.

Our evolutionary tracks were used to estimate stellar masses, ages and local
convective turnover times of stars from the samples of \citet{wright11,wright18}. We
plotted them in the rotation-activity diagram, which presents two main different
regions, saturated and unsaturated. We fitted the data in these two regions with a
two-part power-law function by three different methods. By considering the fit which
reproduces the Sun's position, we found ${\rm Ro_{sat}}$$=$0.087$\pm$0.006,
$\log(L_{\rm X}/L_{\rm bol})_{\rm sat}$$=$$-3.17$$\pm$0.01 and the inclination of the
unsaturated region was found to be $\beta$$=$$-$2.4$\pm$0.1. These are consistent
with others found in the literature and $\beta$ is consistent with a dynamo
efficiency which scales with ${\rm Ro}^{-2}$.  According to our analysis of the
rotation-activity relationship of both partially and fully convective subsamples,
these operate similar dynamos, showing the same dependence of $L_{\rm X}/L_{\rm bol}$
on ${\rm Ro}$. This would imply that the tachocline is not a fundamental ingredient
for the dynamo process.

\section*{Acknowledgements}
The authors thank Drs. Francesca D'Antona (INAF-OAR, Italy) and Italo Mazzitelli
(INAF-IASF, Italy) for granting them full access to the {\ttfamily ATON} evolutionary
code.  We also are grateful to the referee, Dr.\  Christopher Tout, for his many
comments and suggestions that helped to improve this work.  Financial support from
the Brazilian agencies CAPES, CNPq and FAPEMIG is gratefully acknowledged.

\section*{Data Availability}
The authors confirm that the observational data supporting the findings of this study
were obtained by Wright et al. (2011, 2018) and that the theoretical data generated
by this study are available within the article, its supplementary materials or
through requests to the corresponding author.
%
%Datasets for this research are included in Wright et al. (2011, 2018).
%The authors confirm that the data supporting the findings of this study
%are available within the article, its supplementary materials or through requests 
%to the corresponding author. 



%\bibliographystyle{mnras}

%\begin{thebibliography}{99}

%%%%%%%%%%%%%%%%%%%%%%%%%%%%%%%%%%%%%%%%%%%%%%%%%%

%%%%%%%%%%%%%%%%% APPENDICES %%%%%%%%%%%%%%%%%%%%%

%\appendix

% Don't change these lines
\bsp	% typesetting comment
\label{lastpage}
\end{document}